# PAPR Analysis for Dual-Polarization FBMC

Hosseinali Jamal, David W. Matolak
Department of Electrical Engineering
University of South Carolina
Columbia, SC, USA
hjamal@email.sc.edu,
matolak@cec.sc.edu

*Abstract*— In a recent work we proposed a new radio access technique based on filter bank multi-carrier (FBMC) modulation using two orthogonal polarizations: dual-polarization FBMC (DP-FBMC). We showed that with good cross-polarization discrimination (XPD), DP-FBMC solves the intrinsic imaginary interference shortcoming of FBMC without extra processing. DP-FBMC also has other interesting advantages over cyclic prefix orthogonal frequency-division multiplexing (CP-OFDM) and FBMC such as more robustness in dispersive channels, and it is also more robust to receiver carrier frequency offset (CFO) and timing offset (TO). In this paper we analyze the peak to average power ratio (PAPR) of DP-FBMC and compare PAPR simulation results with that of conventional FBMC, for different prototype filters and overlapping factors. According to the analysis and results, with a proper choice of prototype filter, DP-FBMC has comparable PAPR to FBMC.

## I. INTRODUCTION

It is known that orthogonal frequency division multiplexing (OFDM) modulation with a cyclic prefix (CP) extension decreases the spectral efficiency, especially in highly-dispersive channels. Also, because of the rectangular pulse's frequency response with large sidelobes, CP-OFDM thus requires a large number of guard subcarriers to reduce the out-of-band power emission, further decreasing spectral efficiency. As an alternative approach to increase the spectral efficiency, filterbank multi-carrier (FBMC) has been proposed. FBMC does not require CP and has very compact spectral shape due to filtering. Despite these FBMC advantages, it incurs a shortcoming due to "intrinsic imaginary interference" in dispersive channels, therefore it requires extra processing to estimate and mitigate this interference.

In [1], we proposed dual-polarization FBMC (DP-FBMC) which can solve this intrinsic interference problem. Basically by using two polarizations and different multiplexing techniques in FBMC systems we add another dimension to suppress the intrinsic interference. We showed that transmitting symbols on two orthogonal polarizations reduces the interference by a large extent. Using different multiplexing techniques we proposed three different DP-FBMC approaches: *Structure I* uses time-polarization division multiplexing (TPDM), *Structure II* uses frequency-polarization division multiplexing (FPDM), and *Structure III* uses time-frequency-polarization division multiplexing (TFPDM). The difference in these methods is the location of offset-QAM (OQAM) modulated symbols in the time, frequency, and polarization domains. As described in [1], for DP-FBMC *Structure I* we separate the adjacent modulated and filtered OQAM symbols on two orthogonal polarizations by multiplexing symbols in time domain. By this approach we can remove the intrinsic interference that results from (temporally) adjacent symbols. In this paper we focus on *Structure I* as our first DP-FBMC suggestion because this structure has complexity advantage compared to the other two structures. DP-FBMC *Structure I* has similar computational complexity and requires the same equipment and space as conventional FBMC. In [1] via simulation results we showed that in practical cross-polarization discrimination (XPD) conditions in different channel environments the proposed DP-FBMC has similar bit error ratio (BER) performance comparing to conventional FBMC and CP-OFDM, and in highly frequency-selective channels it has better BER performance than FBMC.

In this paper with some analysis and simulation results we investigate and compare the peak to average power ratio (PAPR) of DP-FBMC with that of conventional FBMC and CP-OFDM. In [1] we showed that choosing square-root raised cosine (SRRC) prototype filter for DP-FBMC Structure I we can significantly remove the intrinsic imaginary interference, and in this paper we also show that choosing the same filter with proper length, it can also yield PAPR values nearly identical to that of FBMC and CP-OFDM.

The remainder of this paper is organized as follows: in Section II we review the conventional FBMC and proposed dual polarization FBMC system models. In Section III we provide the PAPR analysis of DP-FBMC. In Section IV we provide simulation results and compare DP-FBMC and conventional FBMC systems' PAPRs for different prototype filters and numbers of subcarrier. In Section V we provide conclusions.

## II. DUAL POLARIZATION FBMC SYSTEM MODEL

In this section first we describe and review the FBMC system based on OFDM-OQAM modulation [2-4]. We then review our proposed DP-FBMC system model [1].

### A. OFDM-OQAM based FBMC

In the OFDM-OQAM structure, we derive the real valued OQAM symbols of subcarrier index *n* and symbol index *m* from complex QAM symbols $d_{n,l}$ as follows, where *l* is the sample index of QAM symbols and *m* is the sample index of modulated OQAM symbols with double rate of QAM samples,

$$a_{n,m} = \begin{cases} \mathcal{R}e(d_{n,l}) & n \text{ even}, l \text{ even}. \\ \mathcal{I}m(d_{n,l}) & n \text{ odd}, l \text{ even}. \\ \mathcal{I}m(d_{n,l}) & n \text{ even}, l \text{ odd}. \\ \mathcal{R}e(d_{n,l}) & n \text{ odd}, l \text{ odd}. \end{cases} \quad (1)$$

According to (1), to obtain the OQAM symbol structure of OFDM-OQAM the real and imaginary components of QAM symbols are offset by a half symbol period. Then in order to achieve the same symbol rate as complex QAM symbols, we double the rate of the OQAM symbols. In next step these OQAM symbols $a_{n,m}$, are filtered through prototype filter $h(t)$ and then modulated across $N$ subcarriers and then phase shifted according to the following continuous form equation,

$$x(t) = \sum_{n=0}^{N-1} \sum_{m \epsilon \mathbb{Z}} a_{n,m} h\left(t - m\frac{T}{2}\right) e^{\frac{j2\pi nt}{T}} e^{j\theta_{n,m}}, \quad (2)$$

where $h(t)$ is the finite impulse response of a filter with length $L=KN$ samples, and $K$ is defined as the overlapping factor. In the OFDM-OQAM structure the phase of the OQAM symbols should be changed according to $\theta_{n,m} = \frac{\pi}{2}(n+m)$ to satisfy the real orthogonality condition at the receiver [2], [3]. Then the filtered symbols are overlapped by half a symbol duration, $T/2$.

The direct form of FBMC modulation (2) has high complexity, therefore in practice similar to OFDM, fast and inverse fast Fourier transforms (FFT, IFFT) and an extra processing block called a polyphase network (PPN) are used (Figure 1). For more details regarding the FBMC PPN structure and FFT implementation refer to [3, 4]. After IFFT processing, the subcarriers will be filtered through the PPN network. The filtered symbols are then overlapped by half symbol period to achieve maximum spectral efficiency. The reverse process will be followed at receiver.

*B. DP-FBMC*

Figure 2 illustrates the idea of transmitted signals on two horizontal and vertical polarization antennas in DP-FBMC *Structure I*. In this Figure and the rest of the paper we use vertical and horizontal polarizations, but we can also choose left and right-handed circular polarizations (or other orthogonal pairs) as dual orthogonal polarizations in DP-FBMC. Therefore, in this paper we use indices *H*, *V* as horizontal and vertical polarizations, respectively.

To better understand the idea of DP-FBMC *Structure I*, in Figure 3 we depict the time-frequency phase-space lattice diagram to illustrate the transmitted OQAM symbols in time, frequency, and phase for an example of 16 subcarriers. This diagram shows how adjacent symbols are separated on two polarizations, also circles and squares indicate the $\pi/2$ phase shift on adjacent symbols to satisfy the real orthogonality [3].

Here we define the polarization multiplexed OQAM symbols in DP-FBMC *Structure I* in (3). Using these multiplexed symbols we can reformulate (2) as in the two equations of (4) for each of the polarizations signals of the DP-FBMC communication system.

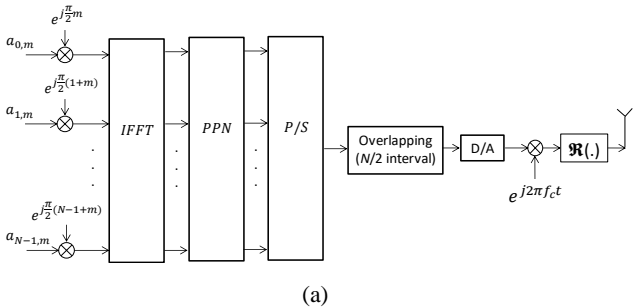

(a)

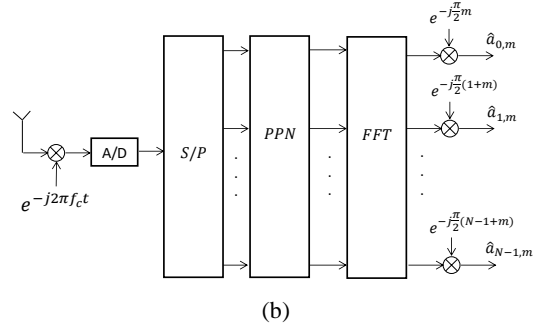

(b)

Figure 1. FBMC communication system block diagram (applicable for DP-FBCM); (a) transmitter, (b) receiver.

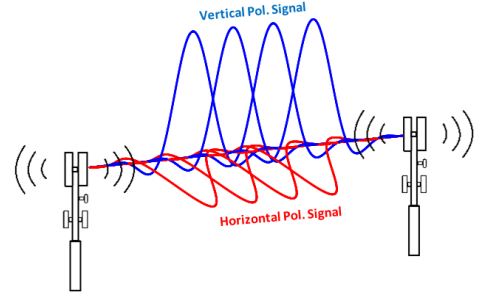

Figure 2. DP-FBMC wireless communication link (*Structure I*).

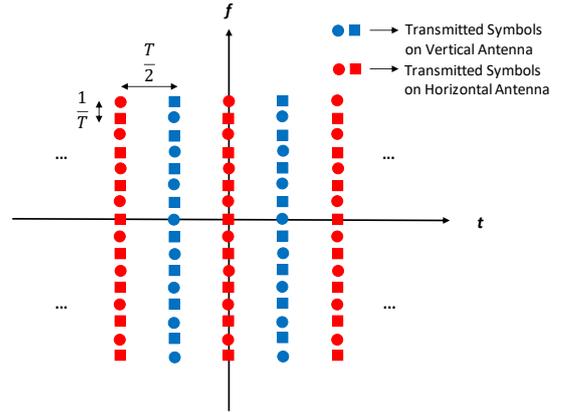

Figure 3. DP-FBMC symbols time-frequency-polarization phase-lattice (*Structure I*).

$$a_{n,m}^H = \begin{cases} a_{n,m} & m \text{ even.} \\ 0 & m \text{ odd.} \end{cases}$$
$$a_{n,m}^V = \begin{cases} a_{n,m} & m \text{ odd.} \\ 0 & m \text{ even.} \end{cases} \quad (3)$$

$$x^H(t) = \sum_{n=0}^{N-1} \sum_{m \epsilon \mathbb{Z}} a_{n,m}^H h\left(t - m\frac{T}{2}\right) e^{\frac{j2\pi nt}{T}} e^{j\theta_{n,m}},$$
$$x^V(t) = \sum_{n=0}^{N-1} \sum_{m \epsilon \mathbb{Z}} a_{n,m}^V h\left(t - m\frac{T}{2}\right) e^{\frac{j2\pi nt}{T}} e^{j\theta_{n,m}}. \quad (4)$$

### III. PAPR ANALYSIS

In this section first we analyze the PAPR of DP-FBMC *Structure I*. Here we note that as long as the statistical characteristics of each polarization's baseband waveform are identical due to this symmetry we only analyze one polarization's waveform (e.g., horizontal polarization). For DP-FBMC we start and follow similar FBMC PAPR analysis approach in [5]. First the baseband equivalent of a discrete-time DP-FBMC signal can be written as follows for $k \in [0, N-1]$,

$$x[k] = \sum_{n=0}^{N-1} \sum_{m \in \mathbb{Z}} a_{n,m}^H h\left[k - m\frac{N}{2}\right] e^{\frac{j2\pi n\left(k-\frac{L}{2}\right)}{N}} e^{j\theta_{n,m}}. \quad (5)$$

Then as explained in [7] the PAPR measure is defined as follows:

$$PAPR = \frac{max_{k \in \{0,\dots,N-1\}} |x[k]|^2}{E\{|x[k]|^2\}}. \quad (6)$$

Here we note that as long as the prototype filter's length is larger than $N$, for a large number of frame symbols, (6) is a good approximation of PAPR similar to OFDM waveform. In general PAPR is a random variable and for convenience we analyze the complementary cumulative distribution function (CCDF), the probability that PAPR exceeds a given value γ. According to (5), at a given discrete time $k$, $x[k]$ is obtained by the summation over the $N$ subcarriers of the following samples,

$$x_n[k] = \sum_{m \in \mathbb{Z}} a_{n,m}^H h\left[k - m\frac{N}{2}\right] e^{\frac{j2\pi n\left(k-\frac{L}{2}\right)}{N}} e^{j\theta_{n,m}}. \quad (7)$$

Assuming $a_{n,m}^H$ are uncorrelated with variance $\sigma_a^2$ and zero mean, then samples in (7) are uncorrelated as well and we have,

$$E\{x_n\} = 0,$$
$$\sigma_x^2 = E\{x_n x_n^*\} = \sigma_a^2 \sum_{m \in \mathbb{Z}} h[k - m\frac{N}{2}]^2. \quad (8)$$

Thus the mean and variance of $x_n[k]$ are independent of $n$. Now from (5) after summing all subcarriers, it is clear that, for $N$ large enough, based on the central limit theorem, $x[k]$ will approach a complex Gaussian distributed process with a zero mean and a variance of $2\sigma_k^2 = N\sigma_x^2$, where $\sigma_k^2$ denotes the variance of the real and imaginary parts of (5), therefore we can say that $|x[k]|$ follows a Rayleigh distribution. Defining $R = |x[k]|^2$, and referring to the analysis in [5] the probability density function of $R$ is derived as follows,

$$p_R(r) = \frac{1}{2\sigma_k^2} e^{-\frac{r}{2\sigma_k^2}}. \quad (9)$$

Now we can define the following for each sample $k$, where $x_0[k]$ is defined as the normalized version of $x[k]$.

$$Y = |x_0[k]|^2 = \frac{|x[k]|^2}{E\{|x[k]|^2\}} = \frac{X}{E\{|x[k]|^2\}}. \quad (10)$$

According to [5], assuming a prototype filter $h$ with unit energy, it can be shown that $E\{|x[k]|^2\} = 2\sigma_a^2$, and then,

$$p_Y(y) = 2\sigma_a^2 p_X(2\sigma_a^2 y) = \alpha_k e^{-\alpha_k y}, \quad (11)$$
where,

$$\alpha_k = \frac{\sigma_a^2}{\sigma_k^2} = \frac{2}{N \sum_{m \in \mathbb{Z}} h[k - m\frac{N}{2}]^2}. \quad (12)$$

Now for a given PAPR value γ, we have,

$$\Pr(|x_0[k]|^2 \leq \gamma) = \int_0^\gamma p_Y(y) dy = 1 - e^{-\alpha_k \gamma}. \quad (13)$$

Assuming that samples of $|x_0[k]|^2$ are independent, then the cumulative density function can be written as,

$$\Pr(PAPR \leq \gamma) = \Pr\left(\bigcap_{i=0}^{N-1} |x_0[i]|^2 \leq \gamma\right)$$
$$= \prod_{i=0}^{N-1} \Pr(|x_0[i]|^2 \leq \gamma) = \prod_{i=0}^{N-1} (1 - e^{-\alpha_i \gamma}). \quad (14)$$

Finally, for the complementary function, we have the following CCDF expression,

$$\Pr(PAPR \geq \gamma) = 1 - \prod_{i=0}^{N-1} (1 - e^{-\alpha_i \gamma}). \quad (15)$$

According to (12) and (15), for every PAPR value, the CCDF is related to the number of subcarriers $N$ and prototype filter characteristics. For FBMC, in [5] it is shown that for square-root raised cosine (SRRC) filters with small enough roll-off factors we have nearly identical PAPR to OFDM. In DP-FBMC using the same SRRC filter we have a similar situation, but from (3) we realize that as long as the multiplexed polarized OQAM symbols are zero for even values of $m$, then according to (8), (12), and (15) we should expect different PAPR results. In DP-FBMC *Structure I*, as will be shown in PAPR results because of time-division multiplexing (TDM) nature and temporal gaps in the waveform, PAPR increases comparing to the FBMC. But we will show that using the SRRC filter with the suggested roll-off factor and increasing overlapping factor we can approach to the FBMC PAPR result.

## IV. SIMULATION RESULTS

In this section first using the suggested square-root raised cosine (SRRC) and another well-known filter used in PHYDYAS [6] project, and different $K$-factors we plot the simulated PAPR results of DP-FBMC *Structure I*, FBMC and CP-OFDM. In Figure 4 the impulse responses of prototype filters for $K$=4 are shown.

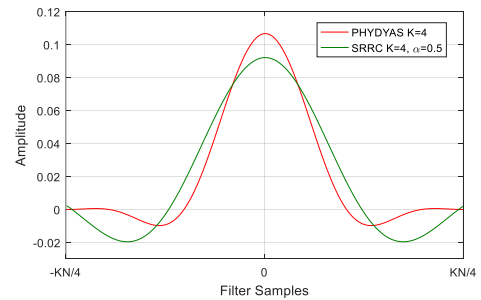

Figure 4. SRRC and PHYDYAS prototype filter impulse responses.

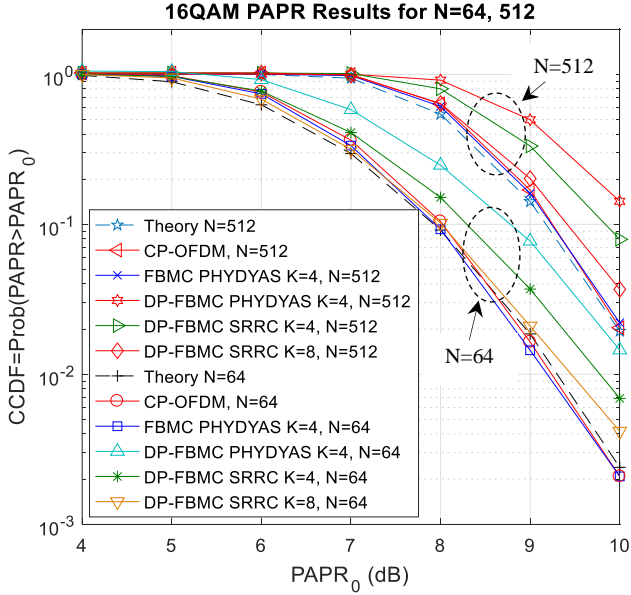

Figure 5. CCDF vs. PAPR$_0$ for different prototype filters (without DFT spreading), 16QAM and $N$=64, 512.

For SRRC in [1] we showed that the roll-off factor equal $2/K$ will result the best BER performance comparing to all prototype filters and we suggested this filter and roll-off factor for DP-FBMC. In the simulation results in this paper we also use the same filter and roll-off factor for comparing PAPR results.

The CCDF results of a vector of logarithmic PAPR values of PAPR$_0$=$10\log(\gamma)$ in the range of 4 to 10 dB for two values of the number of subcarriers $N$ = 64 and 512, are shown in Figure 5. In this Figure we also plot the PAPR approximation result based on (15). As can be seen, DP-FBMC structure $I$ has slightly larger PAPR than FBMC. We note that using SRRC filter with suggested roll-off factor in DP-FBCM *Structure I* comparing to PHYDYAS filter has the advantage of improving the PAPR, and this is because of the larger side lobes of the SRRC impulse response in the time domain (Figure 4). Also similar to BER improvement shown in [1], the suggested SRRC filter and larger $K$ overlapping factors also improve the PAPR results.

In Figure 6, in order to better understand and show the effect of the filter on DP-FBMC waveform, we plot the waveform of one frame of DP-FBMC consisting of 32 symbols and 512 subcarriers (without tails) using SRRC and PHYDYAS filters and identical overlapping factor $K$=4.

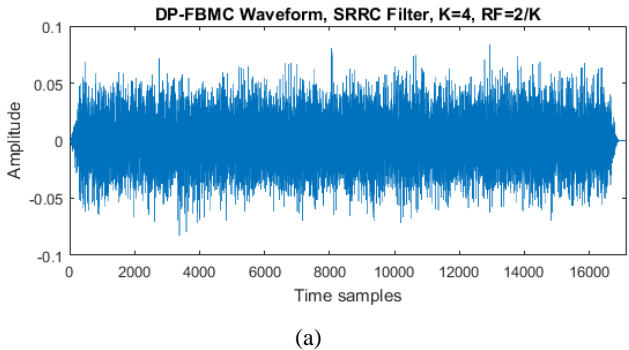

(a)

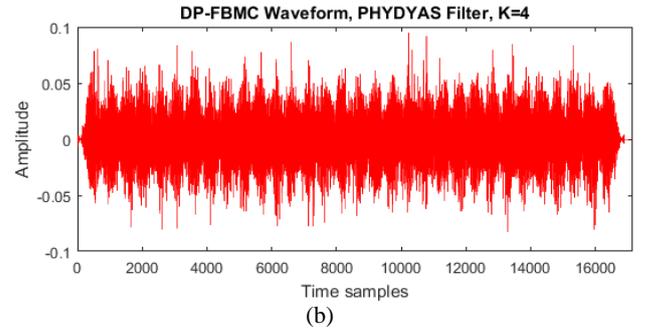

(b)

Figure 6. A frame of DP-FBMC (without tails), 32 symbols, $N$=512; (a) SRRC filter, $K$=4, $\alpha$=0.5, (b) PHYDYAS filter, $K$=4.

As can be seen using the PHYDYAS filter yields temporal fluctuations at every symbol period which affect the PAPR. Here we note that according to our other results (not shown here) *Structures II* and *III* have PAPR similar to that of conventional FBMC and OFDM, and only *Structure I* has worse PAPR performance due to its TDM nature.

In Figure 7 we also compare the PSD of these systems obtained via the periodogram technique. For DP-FBMC waveforms we plot the simulation results before and after removing the two ends of the frames for the FBMC and DP-FBMC waveforms (resulting from filter tails). Thus we truncated the first $(K/2-1)N$ and last $(K/2-1)N$ samples of each frame on DP-FBMC waveforms (as shown in Figure 6). In Figure 7(a) we also plot the PSD of CP-OFDM with and without windowing for comparison. In CP-OFDM windowing is used to reduce the out of band power. As expected, longer SRRC filters (bigger $K$) also yields smaller out of band power. In Figure 7(b) we show the PSDs without truncation for comparison, and as expected the PHYDYAS filter has the most compact PSD.

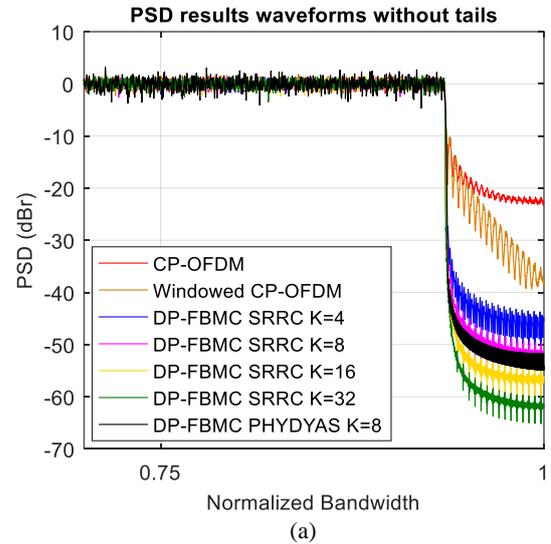

(a)

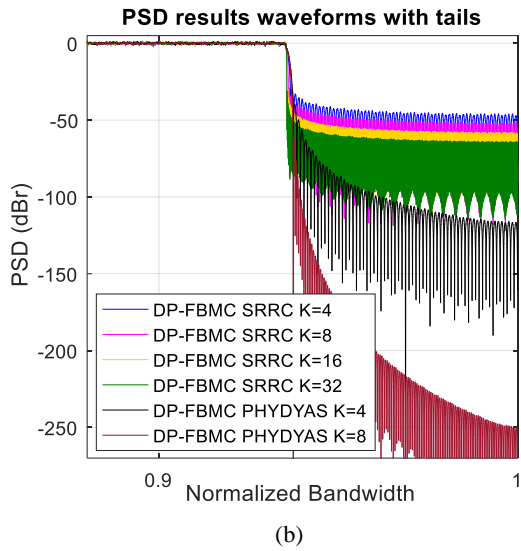

Figure 7. PSD vs. normalized bandwidth; (a) waveforms without tails, (b) waveforms with tails.

## V. CONCLUSION

In this paper we investigated the PAPR of the proposed DP-FBMC *Structure I* communication system. Via simulation results we showed that choosing proper prototype filters with appropriate length, DP-FBMC yields comparable PAPR results to that of FBMC and CP-OFDM.